\documentstyle[12pt,eclepsf]{article}
\newcommand{\postscript}[1]{\hbox{\epsfile{file=#1}}}
\begin{document}
\begin{center}
\begin{Large}
{\bf Parallel computation using}\\
       {\bf generalized models of exactly solvable chaos}
       \footnote{to be published in RIKEN Review No.14(1996)}\\

\end{Large}
\end{center}
\begin{center}
    Ken Umeno\\
   {\it Laboratory for Information Representation}\\
   {\it Research on Brain Information Processing}\\
   {\it Frontier Research Program}\\ 
   {\it The Institute of Physical and Chemical Research (RIKEN)}\\
   2-1 Hirosawa, Wako-shi, Saitama 351-01, Japan  
\end{center}
\par
\vspace{0.3 cm}
\begin{small}
How chaos is useful in the brain information processing  is greatly unknown.
Here, we show that the statistical property of chaos such as invariant measures 
  naturally organized under 
 a great number of iterations of  chaotic mappings
 can be used for some complex computations, while the precise information 
 of initial conditions which vanishes 
in the course of iterations 
deos not matter for this kind of  computations. 
The key observation of the present study is that  
computation using ergordicity of dynamical systems  
  can be thought of as  massively parallel Monte Carlo simulations.  
Here, to avoid  difficulty in elucidating the ergordicity of dynamical systems, 
 we  propose computational schemes using    
 the  generalized class of one-dimensional
  chaos with explicit invariant measures. 
The validity of our results which connect chaos with parallel computation 
   is  checked by the precision 
    computations of some   
   transcendental numbers like \(\pi\). 
\end{small}
\newpage 
\normalsize
One of the most important theoretical challenges in our brain information processing is how 
 a great number of noisy neurons  realize the parallelism 
  of computation in the brain.   
Thus, in relation to this problem, it is of great interest to consider  
whether or how a kind of chaotic dynamics can be harnessed in brain information 
  processing mechanisms.     
However, most studies concerning chaos in the brain are based on the 
empirical computer simulations  
on some specific conditions.
The purpose of the present paper is to show that we can have some ideal 
model of parallel computation using chaos by purely analytical treatment.\\ 
  
 Let us consider the following deterministic process
   \(x_{n+1}=F(x_{n})\),
 where
\(x_{n}\in \mbox{\boldmath$M$}\).
We can say that this dynamical system has the ergordicity if the following relation 
holds:
\begin{equation}
\label{eq:ergordicity}
  \lim_{N\rightarrow \infty}\frac{1}{N}\sum_{i=1}^{N}Q(x_{i})=
 \int_{\mbox{\boldmath$M$}}Q(x)\mu (dx),
\end{equation}
 where \(\mu(dx)\) is the invariant measures on 
 \(\mbox{\boldmath$M$}\) and 
\(Q(x)\) is a regular function which makes R.H.S. of Eq. (\ref{eq:ergordicity}) finite. 
Essentially, 
Monte Carlo simulations can be regarded as 
 evaluating R.H.S. of Eq. (\ref{eq:ergordicity})
by approximately calculating L.H.S. of Eq. (\ref{eq:ergordicity}) under 
the assumption of the ergordicity. However, it is difficult to get  an explicit 
invariant measure 
\(\mu(dx)\) for a given dynamical system and in general, we cannot tell whether the ergordicity 
holds or not for the system.  Without explicit invariant measures, it is meaningless to consider 
the Monte Carlo type computations because  
the integral as  R.H.S. of Eq. (\ref{eq:ergordicity}) is unclear.\\ 

Quite recently, we discovered the two-parameters class of chaotic dynamical systems with 
explicit invariant measures\cite{ku1,ku3}. 
We call this class {\it exactly solvable chaos} because 
we can obtain the 
invariant measures  as well as the general solutions  analytically.
Here, let us briefly review the models of exactly solvable chaos on the unit 
interval  
  \(\mbox{\boldmath$M$}=\left[0,1\right]\equiv \mbox{\boldmath$I$}\).
The first example of exactly solvable chaos was given by Ulam and Von Neumann 
in 1947 \cite{ulam} as follows:
\begin{equation}
\label{eq:ulamn}
   x_{n+1}=F(x_{n})=4x_{n}(1-x_{n}),
\end{equation}
 where the invariant measure has an explicit expression
\begin{equation}
\mu (dx)= \rho(x)dx=
   \frac{dx}{\pi \sqrt{x(1-x)}}.
\end{equation}
Next, in 1985, Katsura and Fukuda\cite{katsura} generalized 
the Ulam=Neumann map  to the following model: 
\begin{equation}
 F(x)=\frac{4x(1-x)(1-k^{2}x)}{(1-k^{2}x^{2})^{2}}
\end{equation}
 for \(0\leq k< 1\).
Its invariant measure can be calculated \cite{ku1} as    
\begin{equation}
\mu(dx)= \rho (x)dx= \frac{dx}{2K(k)\sqrt{x(1-x)
(1-k^{2}x)}},
\end{equation}
where \(K(k)\) is given by the complete elliptic 
 integral
\begin{equation}
K(k)=\int_{0}^{1}\frac{du}{\sqrt{(1-u^{2})(1-k^{2}u^{2})}}.
\end{equation}  
The Katsura=Fukuda map for  \(k=0\) corresponds to  the  Ulam=Neumann map.
Furthermore, the present author generalized 
the Katsura=Fukuda to the following model
\cite{ku3}:
\begin{equation}
\label{eq:tokeru}
 F(x)=\frac{4x(1-x)(1-lx)(1-mx)}{1-2(l+m+lm)x^{2}+8lmx^{3}+
  (l^{2}+m^{2}-2lm-2l^{2}m-2lm^{2}+l^{2}m^{2})x^{4}}
\end{equation} 
with two parameters \(m\) and \(l\) such that  
\(-\infty <m\leq l < 1\). 
If we set \(m=0\), we obtain the Katsura=Fukuda map.
The explicit invariant measure of the generalized  map has the form\cite{ku1} of   
\begin{equation}
\label{eq:gexplicit}
 \mu(dx)= \rho (x)dx=  \frac{dx}{2K(l,m)\sqrt{x(1-x)
(1-lx)(1-mx)}},
\end{equation}
where  
\(K(l,m)\) is given by the integral
\begin{equation}
\label{eq:keisuu}
K(l,m)=\int_{0}^{1}\frac{du}{\sqrt{(1-u^{2})(1-lu^{2})
    (1-mu^{2})}}.
\end{equation}
Here, we call (\ref{eq:tokeru}) {\it the generalized Ulam=Neumann map}.
Furthermore, it is shown  \cite{ku3} that the generalized Ulam=Neumann map (\ref{eq:tokeru}) 
 is the most generalized model of exactly solvable chaos on \(\mbox{\boldmath$I$}\), whose  
  Lyapunov exponent is equal to  \(\log 2\),i.e., we cannot further generalize the map 
(\ref{eq:tokeru}). Thus, we can say that the class of the 
generalized Ulam=Neumann maps is universal in representing  
 exactly solvable chaos on \(\mbox{\boldmath$I$}\). 
We note here that their 
dynamical zeta functions of the generalized Ulam=Neumann maps is the same as the one of the 
original Ulam=Neumann map\cite{ku2}, by  which it means that
the statistical characters such as Kolmogorov=Sinai entropy, 
Lyapunov exponent, and topological entropy do not depend on the parameters \(l\) 
and \(m\).  
Figure 1 shows the shapes of the generalized Ulam=Neumann maps with different parameters \(m\) 
for \(l=0.5\). Interestingly, their shapes for \(m=-20,\cdots,-100\) 
are very similar to the Poincar\'e plots of 
 Belousov=Zhabotinski chemical reaction.\\ 

Using the explicit expression of the invariant measures 
\(\rho (x)\)  
of the generalized  
Ulam=Neumann maps in Eq. (\ref{eq:gexplicit}), 
we obtain the following formula:   
\begin{equation}
\label{eq:parallel}
  \lim_{N\rightarrow \infty}\frac{1}{N}\sum_{i=1}^{N}G(x_{n})
  =\int_{0}^{1}G(x)\rho(x)dx=\frac{1}{2K(l,m)},
\end{equation}  
where \(G(x)=\sqrt{x(1-x)(1-lx)(1-mx)}\).
\(K(l,m)\) is known to be a transcendental real number like \(\pi\).
Let us consider the problem of parallel computing of 
 \(K(l,m)\) using this model of exactly solvable chaos. 
We consider \(P\) independent processors labeled by  \(j(1\leq j\leq P)\). 
First, we give  
 an initial condition \(x_{1,j}\in \mbox{\boldmath$I$}\) 
to each processor. The initial 
conditions are different each other. Then each processor 
 iterates the map (\ref{eq:tokeru}) as \(x_{n+1,j}=F(x_{n,j})\)
to compute the number 
  \(\frac{1}{2\hat K(x_{1,j})}\equiv\frac{1}{N}\sum_{i=1}^{N}G(x_{n,j})\).
Finally, after averaging the above numbers, we have an estimated value of 
\(K(l,m)\) as 
\begin{equation}
     2\tilde K(l,m)\equiv \frac{1}{P}\sum_{j=1}^{P} 2\hat K(x_{1,j}) 
    \approx 2K(l,m). 
\end{equation}
To confirm the validity of this proposed parallel 
 algorithm, we computed \(2\tilde K(l,m)\) for \((l,m)=(0,0),(0.5,-100)\)  and 
\((0.5,-200)\). 
 We set that \(P=100\), \(N=10^{6}\) and 
  \(x_{1,j}=0.0073333371j\) for \(j=1,\cdots,100\).    
\begin{center}
{\bf Table 1: Results for \(P=100\) and \(N=1000000\)}\\[0.2cm]
\begin{tabular}{|l|r|r|r|}
\hline
     & \((l,m)=(0,0)\) & \((l,m)=(0.5,-100)\) & \((l,m)=(0.5,-200)\) \\ \hline
\(2\tilde K(l,m)\) &  3.14167445904031 &
 0.804839363844501 &  0.618717074365706 \\ \hline
  \(2K(l,m)\) &  3.14159265358979 &
 0.804826090564245
  &  0.618817767268870 \\
\hline
\end{tabular}
\end{center}

Table 1 shows that  our numerical estimated values \(2\tilde K(l,m)\) using 
the standard double precision representations of real numbers 
neared the exact values 
\(2K(l,m)\) as  \(|2\tilde K(l,m)-2K(l,m)|<0.0002\).\\ 
 Furthermore, as is shown in Fig.2, our estimated values \(2\tilde K(l,m)\) for 
\((l,m)=(0.5,0.3)\) using 
exactly solvable chaos converge faster 
 than those 
 using the usual Monte Carlo method to evaluate integrals on \(\mbox{\boldmath$I$}\), 
 although both of the mean square deviations are 
 inversely proportional to the iteration number \(N\).  
This fastness of convergence of our method over the Monte Carlo method is owing to the fact that 
\(G(x)=\sqrt{x(1-x)(1-lx)(1-mx)}<1<\frac{1}{\sqrt{(1-y^{2})(1-ly^{2})(1-my^{2})}}\) for  
\(x,y\in \mbox{\boldmath$I$}\) and \(0\leq m\leq l<1\).\\


  This work was supported by the special researcher's program to promote 
  basic sciences at RIKEN.\\ 
 
\clearpage

\begin{figure}[htb]
\begin{minipage}[t]{8.0cm}
\postscript{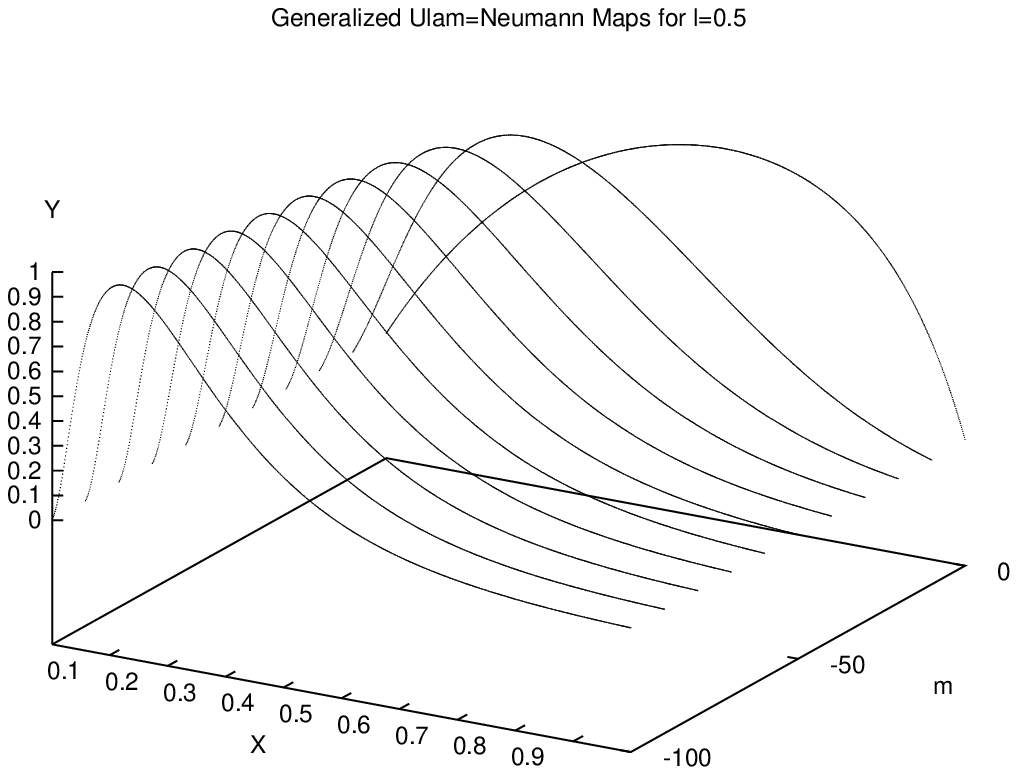,height=10cm}
\end{minipage} \hfill \\
{{\bf Fig. 1:} 
The generalized Ulam=Neumann
maps (7) for \((l,m)=(0.5,0),(0.5,-10),\cdots,(0.5,-100)\).}
\end{figure}
\clearpage 

\begin{figure}[htb]
\begin{minipage}[t]{8.0cm}
\postscript{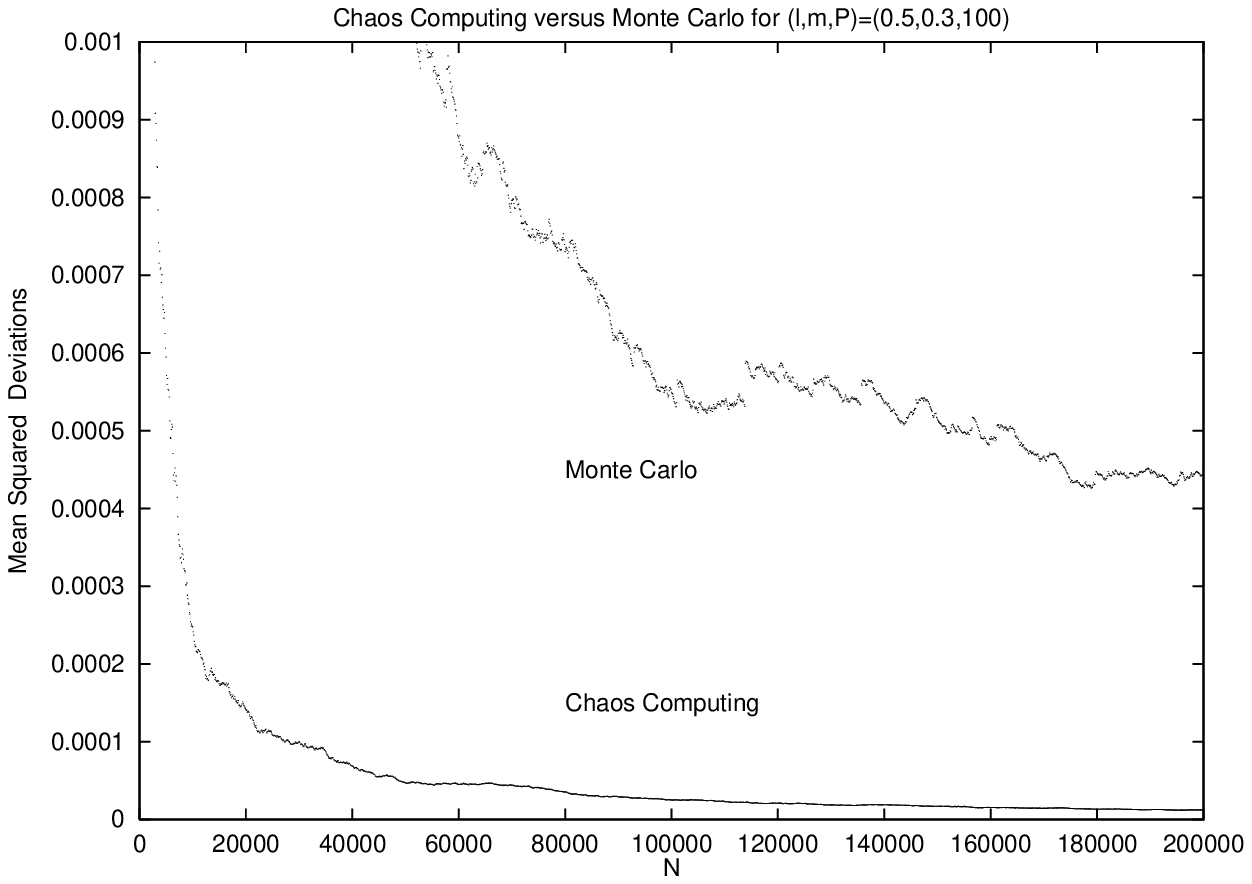,height=10cm}
\end{minipage} \hfill \\
{{\bf Fig. 2:}
Mean squared deviations defined as
\(\frac{1}{P}\sum_{j=1}^{P}\{2\hat K(x_{1,j})-2K(l,m)\}^{2}\) with \((l,m)=(0.
5,0.3)\)
 versus the number of iterations \(N\) for the present method (Chaos Computing)
and for
the usual
method (Monte Carlo), where
the uniform  random numbers were  generated by the function {\it ran(seed=564789
11)} in
the {\it FORTRAN77} library to determine the initial conditions \(x_{1,j}\) for
the
Chaos Computing and
the Monte Carlo paths.}

\end{figure}
   
\end{document}